\documentstyle[12pt,epsf]{article}
\newcommand{\bq}{\begin{equation}}
\newcommand{\eq}{\end{equation}}
\newcommand{\ba}{\begin{array}}
\newcommand{\ea}{\end{array}}
\newcommand{\mathrm}{\rm}

\newcommand {\zf}{$_{Z\!
F}\!I\!^{\textstyle T}\!\!T\!\!_
{{\textstyle E}\!R}$}

\newcommand{\z}{$Z$}

\newcommand{\zp}{$Z'$}

\newcommand{\nobody}{\rule{0ex}{1ex}}
\newcommand{\nn}{\nobody \hfill \\ \noindent }
\begin{document}
  \vspace{-1.4cm}
  \begin{flushright}
  {
  CERN--TH.6545/92 \\}
\vspace*{2.0cm}
\end{flushright}
\vfill

\begin{center}

{\Large \bf
$ZZ'$ Mixing and Radiative Corrections at LEP~I
\vspace*{2cm}\\
}
{\large \bf A. Leike$\nobody^1$,  \,
    S. Riemann$\nobody^1$  \, and \,
    T. Riemann$^{1,2}$ }
\vspace*{1.0cm}\\
$^1${\small DESY --
Institut f\"ur Hochenergiephysik, O-1615 Zeuthen, Germany}
\vspace*{.2cm}\\
$^2${\small Theory Division, CERN, CH-1211 Geneva, Switzerland}
\vspace{1.0cm}\\
\date{\today}

\thispagestyle{empty}
\vfill
{\bf Abstract}
\end{center}
\normalsize
\noindent
We present a method for a common treatment of $Z'$ exchange,
QED corrections, and weak loops in $e^+ e^-$ annihilation.
QED corrections are taken into account by convoluting
a hard-scattering cross section containing $\gamma,\ Z,$ and $Z'$
exchange. Weak corrections and $Z Z'$ mixing are treated simultaneously
by a generalization of     weak form factors.
Using the properly extended Standard Model program for the
$Z$ line shape, \zf,
we perform and compare two different analyses of the 1990 LEP~I data
 in terms of theories
based on the E$_6$-group and in terms of LR-symmetric models.
{}From the LEP~I data alone,
the $Z Z'$~mixing angle may be limited to $|\theta_M| \le 0.01$ and
the $Z'$ mass to $M_2 > 118$--148 GeV,
depending on the model (95\%~CL).

\vspace{2.0cm}
\vspace{2cm}
\begin{flushleft}
{
 CERN--TH. 6545/92  \\
 June 1992      \\}
\end{flushleft}
\newpage
\setcounter{page}{1}
\section{Introduction}
The Standard Model~\cite{no:gws} has been verified with
a precision
including one-loop corrections~\cite{janet}.
Nevertheless, there is a general consensus
that we are
far away from a final understanding of the elementary particle world.
A unification of forces seems to happen at much higher mass scales
than are accessible to present accelerators. Candidates for a
truly unifying theory usually predict additional,
heavy neutral gauge bosons $Z'$
(see e.g.~\cite{673}).

\nn
A search for a $Z'$ at LEP~I energies or below
relies on
minor quantitative modifications
of the neutral current cross sections, and
                  one needs very precise predictions for cross sections
and asymmetries. For a dedicated search, the fermion pair production
reactions at LEP~I are good candidates:
\begin{equation}
e^+e^-  \longrightarrow (\gamma, Z, Z') \longrightarrow f^+f^-(\gamma).
\label{ee}
\end{equation}
A study of these reactions is the subject of the present article.
\\
In principle, the $Z'$ influences cross sections in three different
ways:  \\
$\bullet$
virtual $Z'$ exchange (
also present without $Z Z'$ mixing);
 \\
$\bullet$
shift of the mass of the standard $Z$ boson seen at LEP~I,
due to $Z  Z' \;$ mixing; \\
$\bullet$
modifications of the couplings of the standard $Z$ boson,
due to $Z  Z'$ mixing; this in fact concerns two different, although
related
observables -- the \z $\:$ width {\mbox{[$\sim$ peak height]}} and cross
sections {\mbox{[$\sim$ line shape]}}.
For sufficiently large \zp\ masses,
the direct cross-section contributions originating from
\zp\ exchange may be neglected at LEP~I energies.
On the other hand,
LEP~I {\em is the ideal place to search for the  $Z Z'$ mixing
phenomenon.}

\nn
{}From existing measurements at LEP~I~\cite{bar,zplep,datta},
neutrino physics,
and atomic parity violation~\cite{hollikzp,la2,apv},
it is known that the mixing is very small if not vanishing.
In such a situation, one has to disentangle with great care
both the QED bremsstrahlung and weak standard-theory loop effects
from the \zp\ signals.
Since QED corrections are model-independent (i.e.
well-defined if vector- and axial-vector couplings, mass and width
of the $Z'$ are fixed), the usual convolution  formulae
can be applied
for the total cross section $\sigma_T$ and the forward--backward
asymmetry $A_{FB}$~\cite{npb351}:
\begin{equation}
\sigma_T(s)
 =  \int d v \;
 \sigma_T^{\mathrm{Born}}(s')
 R_T(v),
\label{convol}
\end{equation}
\bq
A_{FB}(s)
 =  \frac{1}{\sigma_T}
 \int d v \; \sigma_{FB}^{\mathrm{Born}}(s') R_{FB}(v),
\label{conafb}
\end{equation}
with $v=1-s'/s$;
the flux factors $R_{T, FB}$ are not influenced by the $Z'$.

\nn
There are two possible approaches to the $Z$ line shape:
 \\
$\bullet$ {\it Indirect data analysis.}
Usually, one unfolds the cross sections and asymmetries
                with some
model-independent ansatz in order to derive e.g. effective couplings
or $Z$ partial widths.      Afterwards, the $Z'$
analysis is performed. 
This seems to be a reliable procedure with the present data, but
may prove to be insufficient in the future.
 \\
$\bullet$ {\it Direct data analysis.}
Alternatively, one can confront~(\ref{convol}) and~(\ref{conafb})
or, equivalently, $ \sigma_{T,FB}^{\mathrm{Born}}(s)$      directly
with the
data. The necessary modifications of these improved Born cross sections
due to         the $Z'$ will be described below.
An  advantage of the        method is the possibility to study
e.g. the top quark and $Z'$ influences on the cross sections
simultaneously. Further,  including the $Z'$ propagator       opens a
window to the $Z'$ mass $M_2$.

\nn
In section~2 we introduce the gauge-boson mixing and define the
notations, while in section~3 the modifications of the weak form factors
due to a $Z Z'$ mixing are explained. Section~4 contains
an application of both analysis methods
to LEP~I data, their comparison,  and a discussion of the perspectives.

\section{Gauge-Boson Mixing}
%
The Lagrangian of the neutral gauge-boson interactions with fermions
\begin{equation}
{\cal L} = eA_\beta J_\gamma^\beta +
g Z_\beta J_Z^\beta + g' Z'_\beta J_{Z'}^\beta
\label{eq31}
\end{equation}
contains currents of the form
\begin{eqnarray}
J_n^\beta = \sum_{f}\;\bar{f} \gamma^\beta\;[v_f(n) + \gamma_5 a_f(n)]\; f,
\hspace{0.5cm} n = \gamma, Z, Z'.
\label{eq32}
\end{eqnarray}
The $Z$-boson couplings are:
\begin{eqnarray}
g = \sqrt{ \sqrt{2} G_{\mu} M_Z^2 }, \hspace{0.5cm}
a_f(Z) \equiv a_f = I_3^L(f),\hspace{0.5cm}
v_f(Z) \equiv v_f = a_f (1-4 |Q_f| \sin^2\theta_W).
\label{smcoup}
\end{eqnarray}
The photon couplings are defined such that $Q_e=-1$.
The couplings $a_f(Z')\equiv a'_f$ and $v_f(Z')\equiv v'_f$ depend on the
particular $Z'$ model. Some popular choices are the $E_6$ model
and the left--right-symmetric model~\cite{673}.
In the following, we will assume that the mass eigenstates
$Z_1$ and $Z_2$ result from a mixing of symmetry eigenstates
$Z$ and $Z'$:
\begin{equation}
\left( \begin{array}{c} Z_1 \\ Z_2 \end{array} \right)
=
\left( \begin{array}{rl}  \cos\theta_M & \sin\theta_M \\
                        - \sin\theta_M & \cos\theta_M \end{array} \right)
\left( \begin{array}{c} Z \\ Z' \end{array} \right)
\label{zzmix}
\end{equation}
In the on-mass-shell renormalization scheme,
the weak mixing angle $\theta_W$ and the gauge-boson mixing angle
$\theta_M$ are related to the gauge-boson masses:
\begin{equation}
\cos\theta_W = \frac{M_W}{M_Z},\hspace{0.5cm}
 \tan^2\theta_M \equiv t_M^2 = \frac{s_M^2}{c_M^2} =
 \frac{M_Z^2-M_1^2}{M_2^2-M_Z^2}.
\label{t2tm}
\end{equation}
Here, $M_W, M_1, M_2$ are particle masses and $M_Z$ has been introduced
for convenience. Without mixing, $M_Z=M_1$. The resonance, which is
being observed at LEP~I, has mass $M_1$ and width $\Gamma_1$.
{}From~(\ref{zzmix}), we deduce the following couplings of
$Z_1$ to fermions:
\begin{eqnarray}
a_f(1) &=&  c_M a_f + \frac{g'}{g} s_M a'_f
       \equiv (1-y_f) a_f,
\\ \nonumber
v_f(1) &=& c_M v_f+\frac{g'}{g}s_M v'_f \equiv
 a_f(1)\left[1+\left( \frac{v_f}{a_f}-1 \right)(1-x_f)\right]
 \nonumber \\
&=& a_f(1)\left[1+4|Q_f|\sin^2\theta_W(1-x_f)\right].
\label{vf1}
\end{eqnarray}
Here the $y_f$ are corrections of the axial couplings and the $x_f$
of the weak mixing angle in the vector couplings.
They are approximately linear in the $Z Z'$ mixing angle:
\begin{eqnarray}
y_f &=& -s_M \frac{g'a_f'}{g a_f}  + (1 - c_M) \sim -s_M \frac{g'a_f'}
{g a_f},
\nonumber \\
x_f &=&      (1-v_f/a_f)^{-1}
\left( \frac{v_f+t_M v'_f g' / g} {a_f+t_M a'_f g' / g} \right)
\sim s_M\frac{g'}{g} \frac{a'_f}{a_f} \frac{v'_f/a'_f-v_f/a_f}{v_f/a_f}.
\label{xf}
\end{eqnarray}
%
 \section{Weak Form Factors}
With a     $ZZ'$ mixing, the matrix element for     reaction~(\ref{ee})
may be written in the form:
\begin{eqnarray}
{\bar{\cal M}}_{1} &\sim&
\frac{1}{s-m_1^2} \frac{G_\mu M_1^2}{\sqrt{2}}
a_e a_f\rho_{ef}^M
\Biggl[ L_\beta \otimes L^\beta
- 4 \, |Q_e |\sin^2 \theta_W \kappa_e^M \gamma_\beta \otimes L^\beta
        \nonumber  \\
& & - 4|Q_f |\sin^2 \theta_W \kappa_f^M L_{\beta} \otimes \gamma^\beta
+ 16 |Q_e Q_f | \sin^4 \theta_W
\kappa_{ef}^M \gamma_\beta\otimes\gamma^\beta \Biggr ].
\label{m1bar}
\end{eqnarray}
The following short notations are used:
\begin{equation}
A_{\beta} \otimes B^{\beta} = \left[ \bar u_e A_{\beta} u_e \right]
                          \cdot \left[ \bar u_f B^{\beta} u_f \right],
\hspace{0.5cm} L_{\beta} = \gamma_{\beta}(1+\gamma_5 ).
\label{lmu}
\end{equation}
In the propagator,
$m_1^2=M_1^2-i s \Gamma_1/M_1$
denotes the complex mass parameter including finite-width
effects\footnote{We do not discuss here problems connected with the
definition of gauge-boson masses depending  on the handling of the
energy dependence of the width.}.
The form factors $\rho_{ef}^M,\kappa_e^M,\kappa_f^M$, and $\kappa_{ef}^M$
are composed of Standard Model weak corrections (contained in
the weak form factors~\cite{akv,15,zfitter}
$\rho_{ef},\kappa_{e},\kappa_f,\kappa_{ef}$) and additional factors due
to gauge-boson mixing:
\begin{eqnarray}
\rho_{ef}^M &=& \rho_{mix} (1-y_e)(1-y_f) \rho_{ef},
\nonumber \\
\kappa_f^M &=& (1-x_f) \kappa_f,
\nonumber \\
\kappa_{ef}^M &=& (1-x_e)(1-x_f) \kappa_{ef}.
\end{eqnarray}
In~(\ref{m1bar}), the coupling constant
$\alpha$
of the on-mass-shell scheme
has been replaced by the muon decay constant:
\begin{equation}
\frac{\pi \alpha}{2 \sin^2 \theta_W \cos^2 \theta_W} =
\rho_{mix} \frac{G_\mu}{\sqrt{2}}
M_1^2 (1 - \Delta r).
\label{gmualf}
\end{equation}
The factor $(1-\Delta r)$ is absorbed in the definition of
$\rho_{ef}$.
The $\rho_{mix}$ was introduced in~(\ref{gmualf}), and consequently
in~(\ref{m1bar}), in order to eliminate $M_Z$ in favour of $M_1$:
\begin{eqnarray}
\rho_{mix} \equiv
\frac{M_Z^2}{M_1^2} =
\frac{1 + t_M^2 \; M_2^2 / M_1^2 }{1 + t_M^2} =
1 + s^2_M \left( \frac{M_2^2}{M_1^2} - 1 \right) =
\frac{M_W^2}{M_1^2 \cos^2 \theta_W}.
\label{rhomix}
\end{eqnarray}
The last one in the above sequence of equations is valid only for
restricted Higgs sectors.
In the general
case,       $\rho_{mix}$ is an additional free parameter
{}~\cite{degrassisirl,jegerl}.

\nn
The four form factors $\rho^M, \kappa^M$
describe the weak radiative corrections completely in the case of
massless fermions;
the Born amplitude is obtained
for $\rho=\kappa=1$.
The form factor
$\rho_{ef}^M$ can be absorbed by the Fermi constant:
\begin{equation}
G_{\mu}  \rightarrow \bar G^M_{\mu} =
\rho_{ef}^M(s, \cos\vartheta; m_t, M_H, M_1; M_2, \theta_M,
\ldots) G_{\mu}.
\label{gmu}
\end{equation}
Similarly, the
form factors $\kappa^M$ can be interpreted as
renormalizations of the weak mixing angle $\sin^2\theta_W$:
\begin{equation}
\sin^2\theta_W  \rightarrow \left\{
                            \begin{array}{l}
                            \kappa^M_e
                            \sin^2 \theta_W
                            \\
                                          \kappa^M_f
                                          \sin^2 \theta_W
                                          \\
                                 \sqrt{\kappa^M_{ef}}
                                 \sin^2 \theta_W.
                            \end{array}
                            \right.
\label{swkap}
\end{equation}
At LEP~I energies,       an effective weak mixing
angle is often used,
\begin{equation}
\sin^2\theta_W^{\mathrm{eff}}= \kappa \; \sin^2 \theta_W,
\end{equation}
where $\kappa$ may be any (real part of) one of the
form factors $\kappa_f^M$, calculated at
$s=M_Z^2$. For further details
see~\cite{zfitter,altarjada,sommi}.

\nn
To complete the discussion of the $Z$-boson matrix element, we must
define 
yet    the decay width,
which is the sum over all open fermion channels
at    the $Z_1$ mass:
\begin{eqnarray}
\Gamma_1 = \sum_f {\bar \Gamma}(1)_f = \sum_f c_f
          \frac{G_{\mu}}{\sqrt{2}} \frac{M_1^3}{6 \pi}
       \left[ {\bar v}_f^\Gamma(1)^2 + {\bar a}_f^\Gamma(1)^2  \right].
\label{zpwidth}
\end{eqnarray}
For the partial widths, the effective couplings
are:
\begin{eqnarray}
\bar a_f^\Gamma(1) &=& \sqrt{\rho_f^{M,Z}} \: I_3^L (f),
\nonumber \\
\bar v_f^\Gamma(1)
&=&\bar a_f^\Gamma(1) \left[ 1 - 4 |Q_f |\sin^2 \theta_W
                                        \kappa^{M,Z}_f \right]  ,
\end{eqnarray}
where again weak corrections and the $Z Z'$ mixing are properly
combined:
\begin{eqnarray}
\rho_f^{M,Z} &=& \rho_{mix} (1-y_f)^2 \rho_f^Z,
\nonumber \\
\kappa_{f}^{M,Z} &=& (1-x_f) \kappa^Z_{f}.
\end{eqnarray}
The $\rho_f^Z, \kappa_f^Z$ are the weak form factors of the Standard
Model~\cite{zfitter,akhund}.
As is well-known, at LEP~I energies
the couplings in the partial widths differ only slightly from
those in the cross sections.

\nn
We      shortly mention      the matrix element ${\bf\cal M}_{2}$
with exchange of the heavy-mass eigenstate~$Z_2$:
\begin{equation}
{\bf\cal M}_2 \sim \frac{g'^2}{s-m_2^2}
 \left\{ \gamma_\beta \left[ a_e(2) \gamma_5 + v_e(2) \right]
  \otimes
   \gamma^\beta \left[ a_f(2) \gamma_5 + v_f(2)\right]
\right\}   ,
\end{equation}
where $v_f(2), a_f(2) $ are vector- and axial-vector
couplings of the       $Z'$.
After adding  up the photon-exchange diagram ${\bf\cal M}_{\gamma}$
with running QED coupling $\alpha(s)$, the net matrix element is
obtained,
\begin{equation}
{\bf\cal M} = {\bf\cal M}_{\gamma} + {\bf\cal M}_{1} +
                {\bf\cal M}_{2},
\end{equation}
and the improved Born cross sections
$\sigma_{T,FB}^{\mathrm{Born}}(s)
  \sim | {\bf\cal M}|^2$
can be calculated and
      convoluted in~(\ref{convol}) and~(\ref{conafb}).

\nn
At the end of this section, we should mention that     the above
derivations of matrix elements and
form factors are equally valid for Bhabha  and $ep$ scattering
Another remark concerns some underlying assumptions,          made
in the
numeric investigations of the next      section,
which are not inherent in the formalism.
Additional degrees of freedom from exotic fermion mixing and Higgs
structures are investigated in detail in~\cite{673,buchmuwi,nardi}
and will be neglected here.
Further,
it has been pointed out in~\cite{nardi}
that including only the Standard Model
radiative corrections (as is done here) is, in fact, a reasonable
approximation to a complete treatment.

\section{Applications and Discussion}
Based on the above considerations, we created a FORTRAN program
{\tt ZEFIT}~\cite{zefit}, which allows,
together with the Standard Model program
\zf~\cite{zfitter}, 
to search for signals from both the $Z'$ propagator
and a $ZZ'$ mixing in $e^+e^-$ annihilation.

\nn
In Fig.~1, the combined effect
of $Z'$ mass and gauge-boson mixing at the $Z$~peak is shown for one of
the E$_6$-based models, the           $\chi$ model with $\theta_E=0$
(which is, at the same time, one of the LR-models with $\alpha_{LR}=
\sqrt{2/3}$).

 \hspace*{-1.cm}
\begin{minipage}[t]{16.0cm}  {
\begin{center}  \mbox{
\epsfxsize=8.0cm
\epsfysize=10.cm
\epsffile{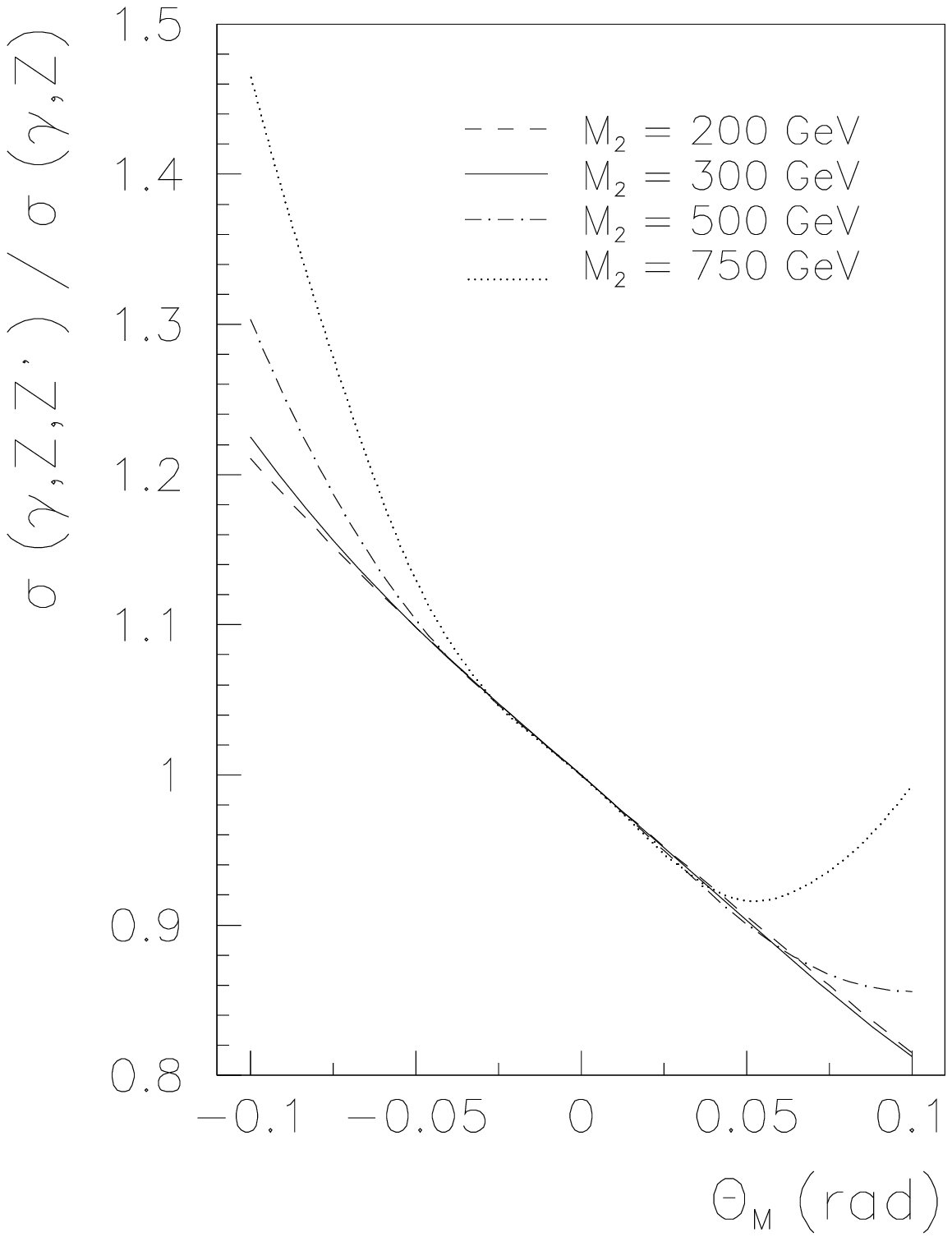} }
   \end{center}
}\end{minipage}
%
\\ \noindent
{\bf Figure 1:} {\it
The ratio
$\sigma^{\mu}_T(\gamma,Z,Z')/\sigma^{\mu}_T(\gamma,Z)$
in the E$_6$-based $\chi$ model as
a function of the $ZZ'$ mixing angle $\theta_M$
at $\sqrt{s}=M_1=91.180$ GeV, $m_t=150$~GeV,
$m_H=300$ GeV. Parameter: the $Z'$ mass $M_2$.
}

\hspace*{-1.cm}
\begin{minipage}[t]{7.8cm} {
 \begin{center}  \mbox{
\epsfxsize=7.8cm
\epsfysize=8.0cm
\epsffile{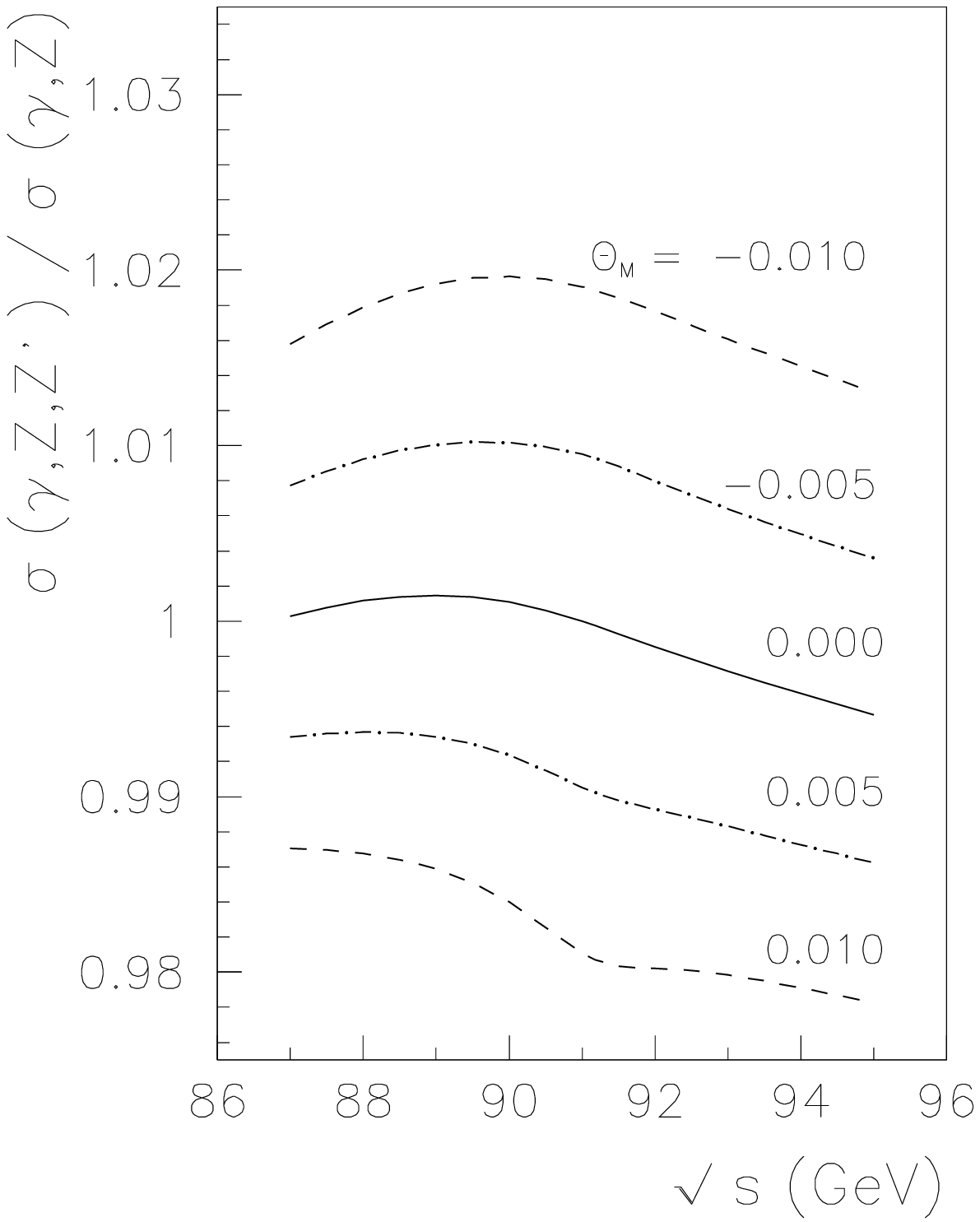}
  }\end{center}
}\end{minipage}
\hspace*{-1.0cm}
\begin{minipage}[t]{7.8cm}{
\begin{center}  \mbox{
\epsfxsize=7.8cm
\epsfysize=8.0cm
\epsffile{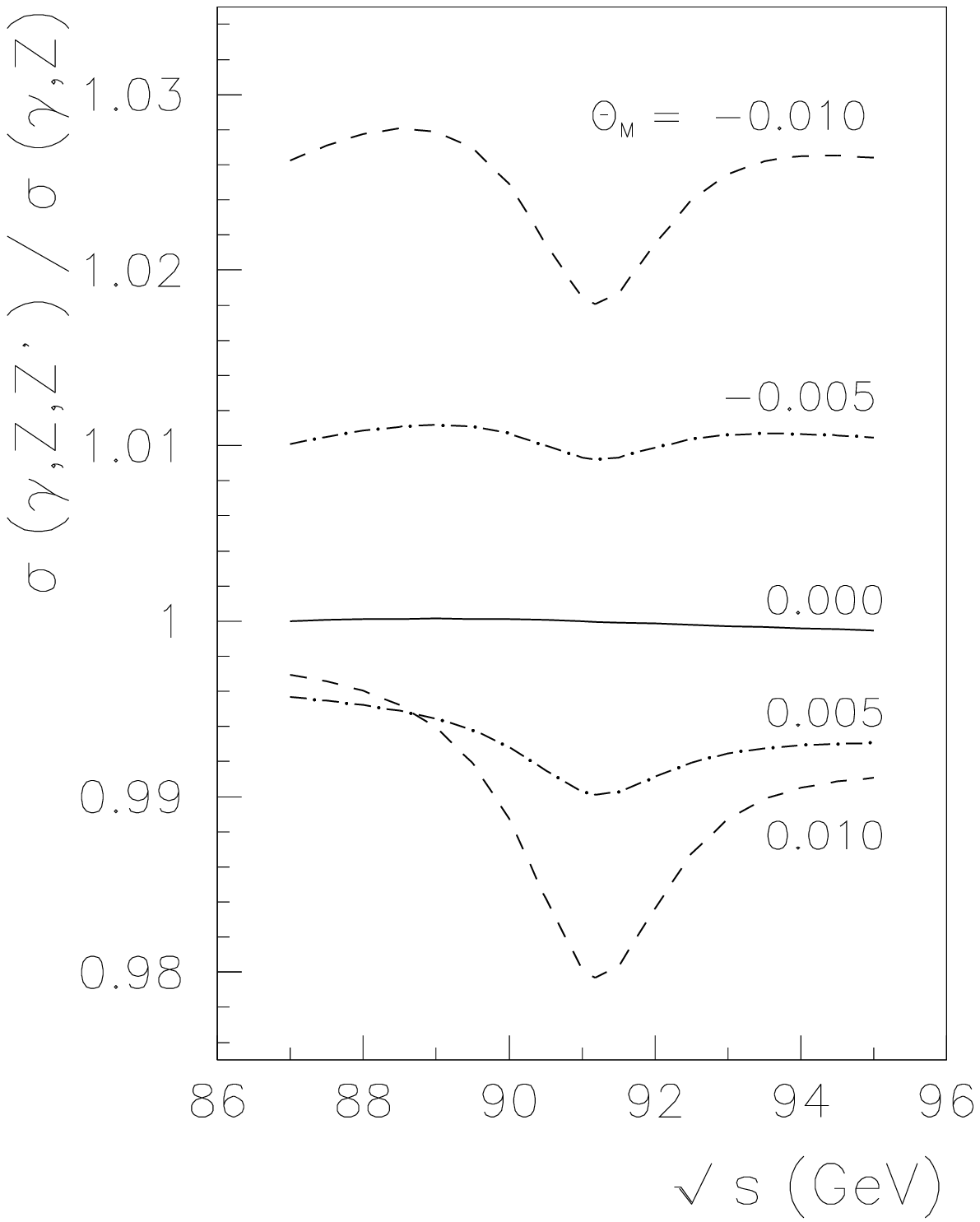}
  }\end{center}
}\end{minipage}

 \nn
{\bf Figure 2:}{  \it
The same as in Fig.1, now as a function of $\sqrt{s}$; $M_2 =250$
GeV (a), 750 GeV (b).
}

\nn
The ratio of muon-production cross
sections $\sigma^{\mu}_T$
with and without $Z'$ is shown as a function of $\theta_{M}$
for different values of the $Z'$ mass.
For
 $\theta_M\le 0.05$,
the ratio is linear in $\theta_M$ and independent of $Z_2$.
This is a consequence of the vanishing $ZZ'$ interference
and of the cancellation of     $\rho_{mix}$ in the
numerator and denominator of the cross-section formula
at $\sqrt{s}=M_1$.
A similar behaviour may be observed for the forward--backward asymmetry.

\nn
For the same model, Figs.~2a and b show this cross-section
ratio as a function of the
centre-of-mass energy for two different $Z'$ mass values.
At the $Z$
peak, the predictions for different values of $M_2$
agree, while they show a different behaviour
off the resonance position.
At extreme LEP~I energies, the differences
reach the order of a percent even for not too large mixing angles.
In view of plans for a high-luminosity version of LEP~\cite{highlumi},
it could be worthwhile to study possible prospects
of this behaviour.

\nn
After these introductory remarks, we now outline the results from
two different $Z'$ search strategies.

\subsection*
{\it 4.1 \hspace*{1.0cm}
Indirect analysis using model-independent parameters}
For our first series of fits we
 used the following input parameters, which we have taken from a
model-independent analysis of 1990~data from all LEP~I collaborations
(Tables~1 and~2 of~\cite{comblep}):
\begin{eqnarray*}
M_1, \, \, \,  \Gamma_1, \, \, \,
\sigma_{\mathrm{had}}^{0,\mathrm{peak}}, \, \, \,
 v_l^2(1), \, \, \,  a_l^2(1),
\end{eqnarray*}
which are
mass and width of the $Z$ boson, the improved  hadronic Born cross
section at the peak,
and the squared effective leptonic couplings to the $Z$-mass eigenstate,
respectively.

\clearpage

\begin{minipage}[t]{7.0cm}{
\begin{center}  \mbox{
\epsfxsize=7.0cm
\epsfysize=7.0cm
\epsffile{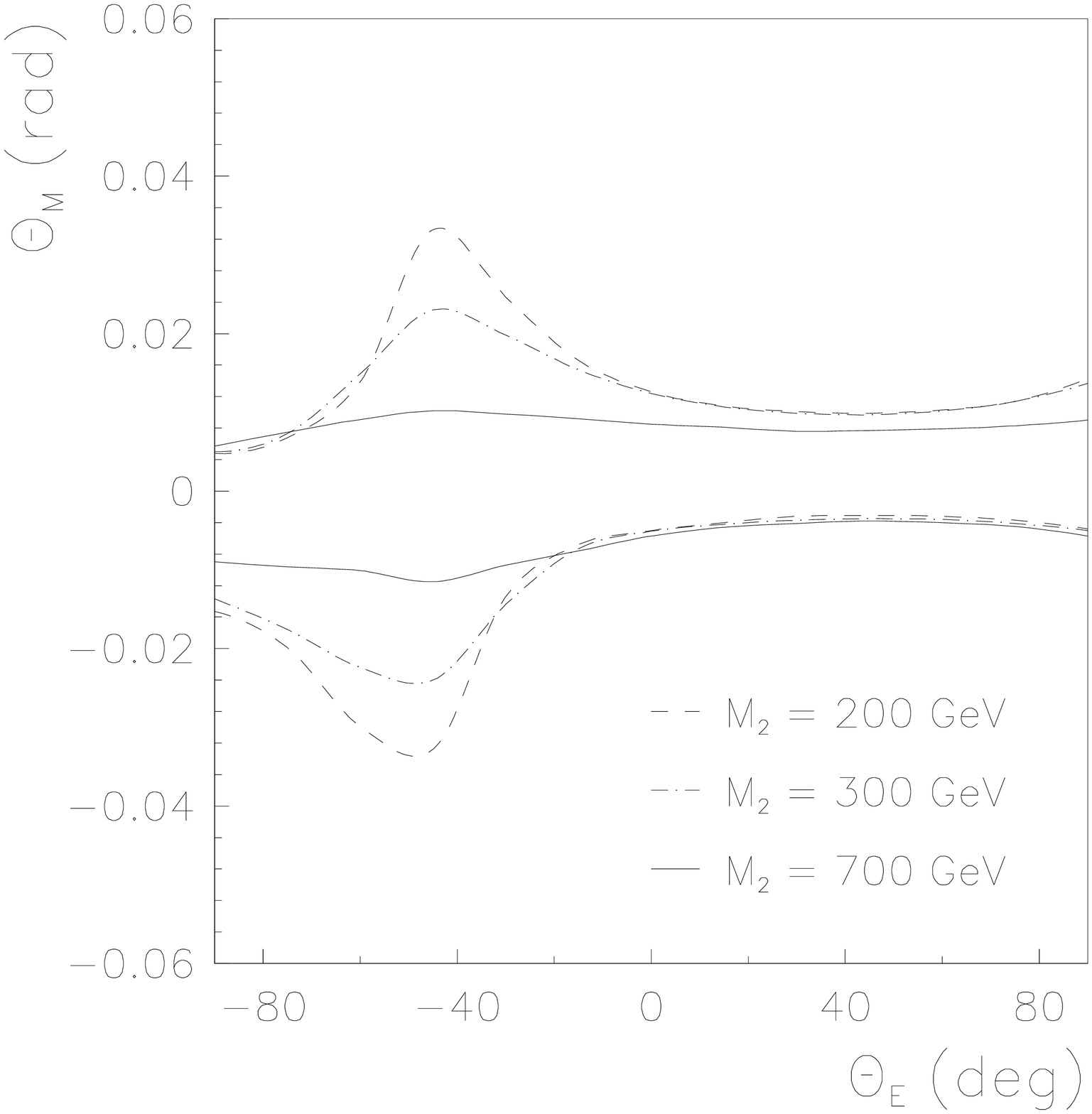}
  }\end{center}
}\end{minipage}
\begin{minipage}[t]{7.0cm}{
\begin{center}  \mbox{
\epsfxsize=7.0cm
\epsfysize=7.0cm
\epsffile{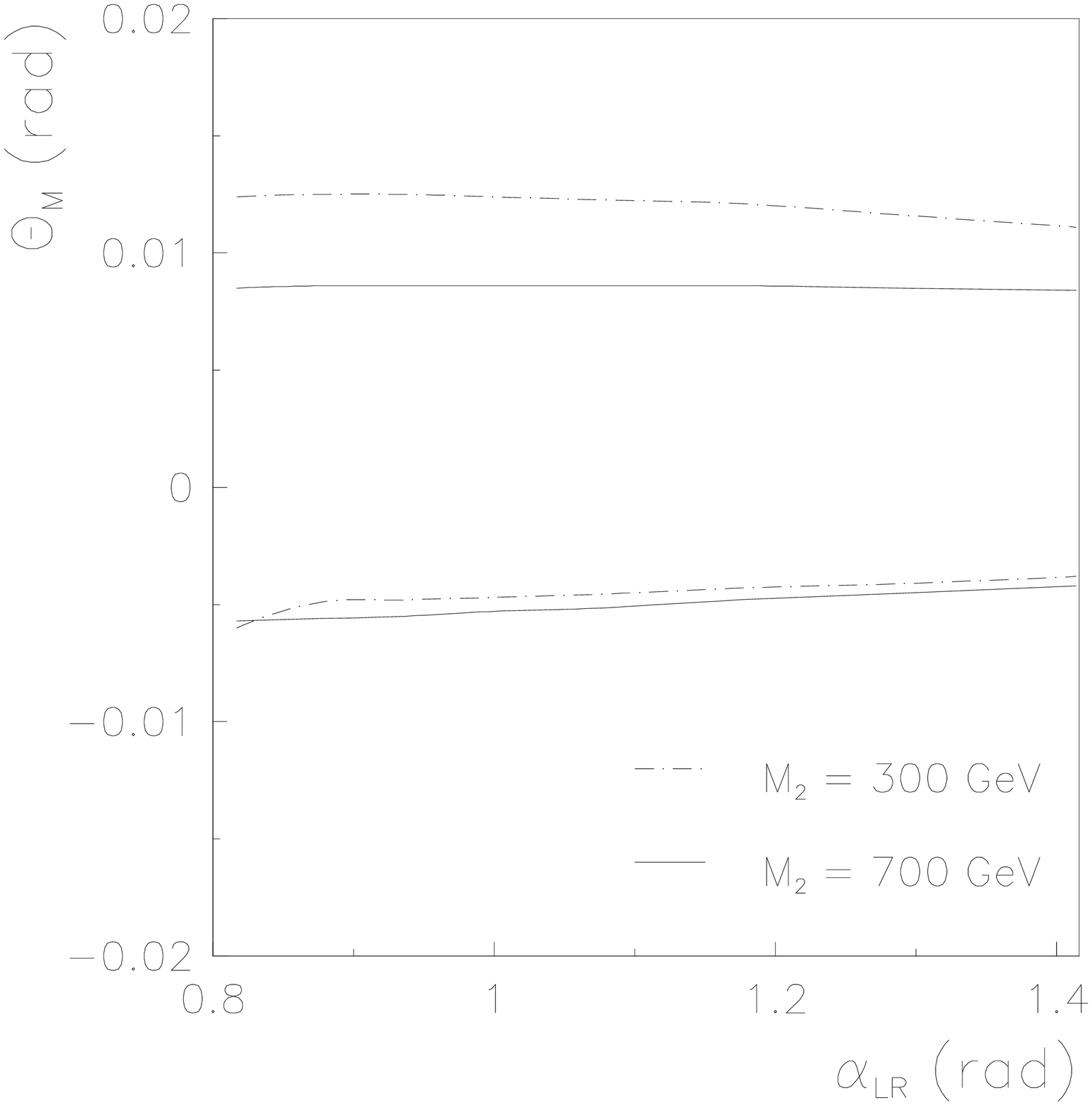}
  }\end{center}
}\end{minipage}
\\ \noindent

\nn
{\bf Figure 3:}{  \it
The 95\%~CL limits for the $ZZ'$ mixing angle $\theta_M$
and $Z'$ mass $M_2$,
derived from a model-independent analysis of LEP~I data
for
two classes of models: (a)~E$_6$-based GUTs, (b)~LR-symmetric theories.
Parameters: $\alpha_s=0.12, m_t=150$ GeV, $M_H=300$ GeV.
}

\nn
Error correlations as given in~\cite{comblep}
are exactly taken into account.
Allowed regions for the $ZZ'$ mixing angle
are shown in Figs.~3a and b for the E$_6$- and LR-models
as functions of their parameters.
The limits depend only weakly on the $Z'$ mass and (not shown here)
on the values of the top-quark mass $m_t$ and strong-interaction
constant $\alpha_s$.

\nn
With Fig.~3,
we obtain limits similar to those of other authors, e.g.~our
Fig.~3a is numerically
comparable with Fig.~2 of~\cite{zplep} where, in a slightly
different approach,                      90\%~CL limits are derived
from the 1990~LEP data;
our Fig.~3b is in         agreement with e.g.
Fig.~3 of~\cite{layssac}.
Both our figures contain slightly better limits than Figs.~3 and~4
of~\cite{renard2}, which summarize an
analysis of the preliminary 1991~LEP data
                             (seemingly 90\%~CL).

\subsection*
{\it 4.2 \hspace*{1.0cm}
Direct analysis of $\sigma_T(s)$ and $A_{FB}(s)$}
Now we discuss  direct fits to cross sections and asymmetries,
taking into account their energy dependence. 
With the rising quality of the data, this 
approach will become more and more advantageous in comparison
to the indirect fits.
An important feature is the immediate use of            line-shape
formulae, including the virtual $Z'$ exchange.                The
influence of   the latter, and    the resulting sensitivity of LEP~I
data to $M_2$ may be    estimated  as follows (similar estimates for
the mixing angle $\theta_M$ are left to the reader):
For sufficiently small $Z Z'$ mixing,
the dominant $Z'$ term at LEP~I is the $Z Z'$
interference.
In a self-explanatory notation,          the line shape is, without
the $Z' $:
\bq
\sigma(s) \sim \frac{r_{\gamma}}{s} +
R \frac{s+R_f(s-M_1^2)} {(s-M_1^2)^2 + M_1^2 \Gamma_1^2} + \ldots,
\label{modint}
\eq
where $R_f=i/R$, and $i$ is the $\gamma Z$~interference.
The $Z Z'$~interference may be interpreted as a
small correction to    the $\gamma Z$ interference~\cite{jegerl}:
\bq
\Delta R_f(Z') \equiv
   R_f' = - 2 \, \frac{g'^2}{g^2} \frac{M_1^2}{M_2^2-M_1^2}
\frac { (v_e v_e' + a_e a_e') \sum_q (v_q v_q' +  a_q a_q') }
      { (v_e^2    + a_e^2   ) \sum_q (v_q^2    + a_q^2    ) }.
\label{rfp}
\eq
With $2 g'^2/ g^2 = (10/3) \sin^2 \theta_W \approx 0.77$~\cite{renorm},
and assuming, for instance,
 for a first estimate, formally $v'=v, a'=a$, this is a rather
simple expression, depending only on the two masses. Further, it is known
how the peak position is shifted by such a $\gamma Z$ interference:
\bq
\Delta \sqrt{s_{\max}} = \frac{1}{4} \frac{\Gamma_1^2}{M_1} R_f'
\approx 17 \, \, \mbox{MeV} \, \, R_f'.
\label{shift}
\eq
A neglect    of this peak shift leads to        a systematic error
 of sign opposite to that
of the $Z$ mass $M_1$. Thus, (\ref{rfp}) and (\ref{shift})
   allow a rough   estimate of the
sensitivity of LEP~I to a $Z'$ propagator;
for instance, with a $\Delta M_1 =
\pm 8 $ MeV, a $Z'$ with a mass of 150 GeV and Standard-Model
couplings cannot be excluded.

\nn
In practice, however, the sensitivity may deviate from   this crude
estimate.
As an example, we use the hadronic line-shape
data and the leptonic line-shape and asymmetry data of the 1990 LEP
runs as quoted in~\cite{comblep}, and references therein,
 for a search of
the allowed region in the $\theta_M$--$M_2$ plane.
The result is shown in Fig.~4 for
   three often analyzed E$_6$-based models ($\theta_{\chi}=0,
\theta_{\psi}=\pi/2, \theta_{\eta}=-52.24^{\circ}=-0.9117$).
The top-quark mass dependence is indicated and, although
present,       not too large.
For the $Z'$ masses,
the (95\%~CL) exclusion limits are:
$M_{\chi}>148$ GeV, $M_{\psi}>122$ GeV, $M_{\eta}>118$ GeV.
In obtaining
these values, we have checked, that the lower $Z'$ mass limits
are
stable against a variation of the $Z$ mass within its experimental
error.
Our   limits are to be compared with the ones derived
in~\cite{barbara} from the CDF search for heavy bosons~\cite{cdf},
$M_2>148, 140, 165$ GeV, respectively, and    similar limits
derived mainly from low-energy physics~\cite{la2}.
Although the present LEP~I $Z'$~mass limits cannot compete with the
world's best estimates, they indicate the potential of this device
if used in the high-luminosity regime.

\nn
Basically, with the exclusion of the low-mass region of the $\eta$ model,
the limits to the $Z Z'$ mixing are nearly independent of   $M_2$.
We should like to compare the allowed regions of the $Z Z'$ mixing
determined in the two approaches.
The limits                   on the $ZZ'$ mixing angle in Fig.~3a
agree perfectly, for the available data,
                                   in their findings
for the $\chi$  and $\psi$ models (Figs.~4a,b).
For the $\eta$ model, there are slight deviations.
For instance, for     $M_2=200$ GeV, one derives from Fig.~3a
$\theta_{\eta} = -0.06 --   0.01$,
 while from Fig.~4c
$\theta_{\eta} = -0.04 --   0.015$. We interpret this as an indication
of the importance of the $Z'$ propagator and of the correct
energy dependences in general for the results in this parameter region.

\begin{minipage}[t]{7.8cm}{
\begin{center}  \mbox{
\epsfxsize=7.8cm
\epsfysize=9.5cm
\epsffile{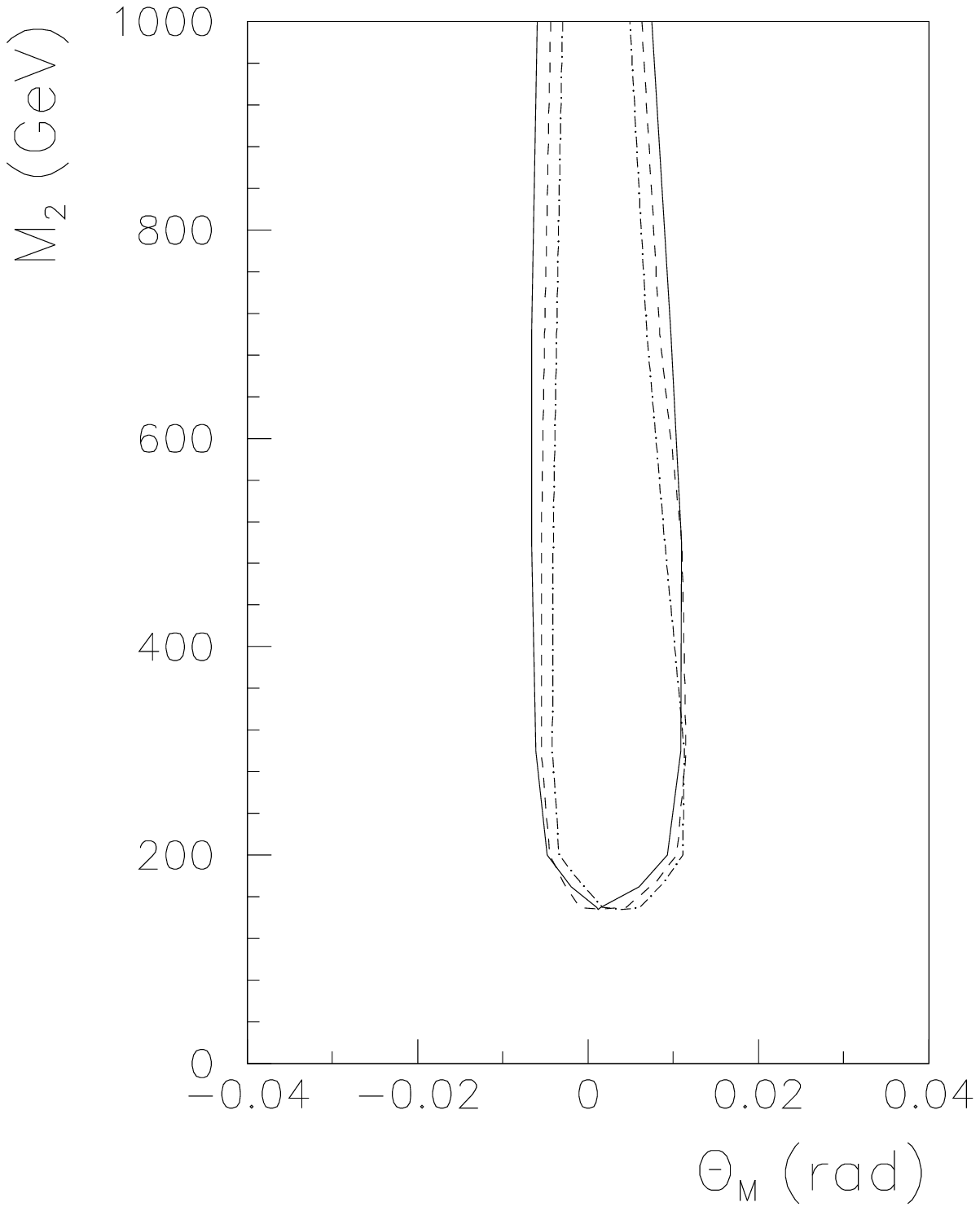}
  }\end{center}
}\end{minipage}
\begin{minipage}[t]{7.8cm} {
\begin{center}  \mbox{
\epsfxsize=7.8cm
\epsfysize=9.5cm
\epsffile{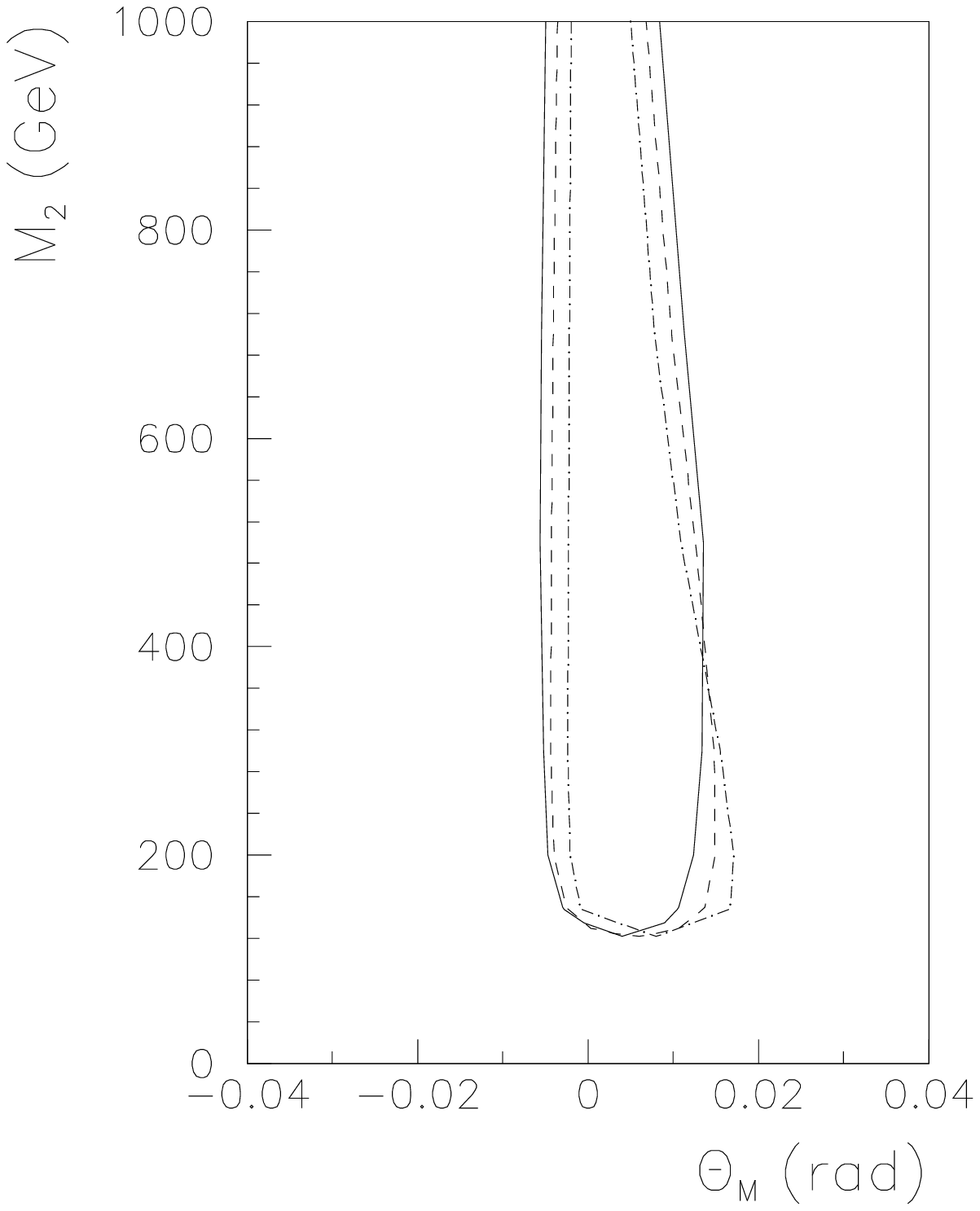}
  }\end{center}
}\end{minipage}

\begin{minipage}[t]{7.8cm}  {
\begin{center}  \mbox{
\epsfxsize=7.8cm
\epsfysize=9.5cm
\epsffile{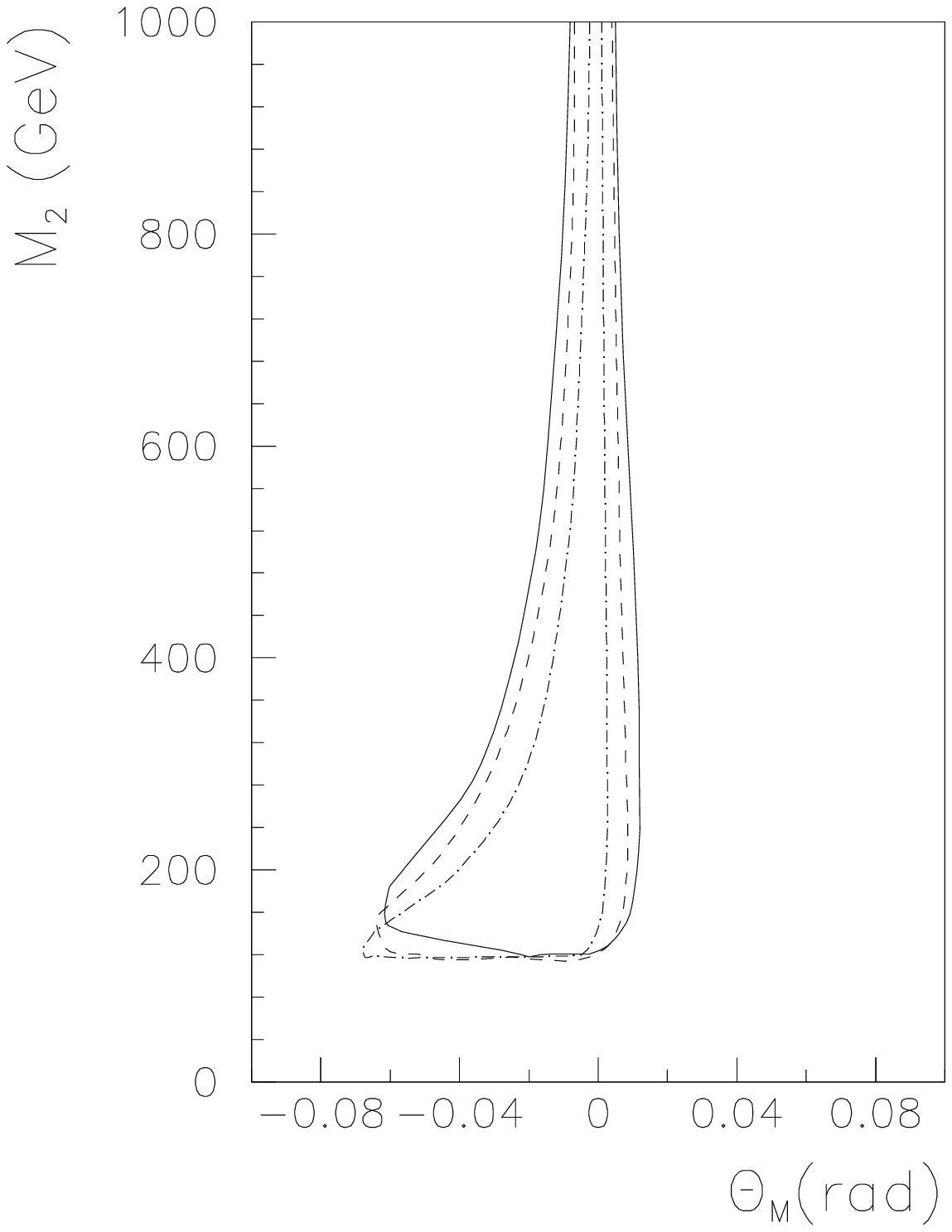} }
   \end{center}
}\end{minipage}
\\ \noindent
{\bf Figure 4:} {\it
Regions of $\theta_M$ and $M_2$ values in the E$_6$-based models
$\chi, \psi, \eta$, which are compatible with the
1991 LEP~I data (95\%~CL). Parameters are $\alpha_s=0.12$, $M_H=300$
GeV; $m_t = 100, 150, 200$~GeV (solid, dashed, dash-dotted curves).
}

\clearpage
{\em To summarize},
we developed two descriptions of fermion pair
production at LEP~I for $Z'$ models, one of them including
the $Z'$ propagator and
$ZZ'$ mixing together with weak corrections and QED corrections.
Some typical applications have been performed with data from the 1991
LEP~I running periods.
Both a fit to model-independent parameters and a direct
line-shape analysis have been performed; they agree for most of the
mixing-angle
limits with each other and with earlier determinations.
Additionally, from the direct fit one may determine
                           $Z'$ mass limits.
Future applications have been indicated.

\subsection*{Acknowledgements}
We would like to thank A.~B\"ohm, S.~Ganguli, D.~Schaile and
C.~Verzegnassi for discussions and valuable hints.
%


\begin{thebibliography}{99}
\bibitem{no:gws}
    S.L. Glashow, {\it Nucl. Phys.} {\bf 22} (1961) 579;

    S. Weinberg, {\it Phys. Rev. Letters} {\bf 19} (1967) 1264;

    A.  Salam, in: N.~Svartholm (ed.),
    {\it  Elementary Particle Theory },
    Stockholm (1968), p.~367.
\bibitem{janet}
J. Carter, Precision Tests of the Standard Model at LEP,
in: S. Hegarty et al. (eds.), {\it
Proc. LP--HEP~91 Conference}, Geneva, 1991
(World Scientific, Singapore, 1992), Vol.~2, p.~3.
\bibitem{673}
J.L. Hewett and T.G. Rizzo,
{\it Phys. Reports} {\bf 183} (1989) 193; \\
P. Langacker, M. Luo and  A. K. Mann,
{\it Rev. Mod. Phys.} {\bf 64} (1992) 87,
and references therein.
\bibitem{bar}
V. Barger, J.L. Hewett and  T.G. Rizzo,
{\it Phys. Rev.} {\bf D42} (1990) 152.
\bibitem{zplep}
G. Altarelli et al., 
{\it Phys. Letters} {\bf B261} (1991) 146; {\bf B263} (1991) 459.
\bibitem{datta}
G. Bhattacharyya, A. Datta, S. N. Ganguli and A. Raychaudhuri,
{\it Mod. Phys. Letters} {\bf A6} (1991) 2551.
\bibitem{hollikzp}
F. del Aguila, W. Hollik, J. M. Moreno and M. Quiros,
{\it Nucl. Phys.} {\bf B372} (1992) 1.
\bibitem{la2}
M.C. Gonzalez--Garcia and J.W.F. Valle,
{\it Phys. Letters} {\bf B259} (1991) 365;
\\
P. Langacker and M. Luo, {\it Phys. Rev.} {\bf D45} (1992) 365.
\bibitem{apv}
K. Mahanthappa and P. Mohapatra, {\it Phys. Rev.} {\bf D43}
  (1991) 3093;\\
P. Langacker, Phys. Letters {\bf B256} (1991) 277.
\bibitem{npb351}
    D. Bardin et al.,
    {\it Nucl. Phys.} {\bf B351} (1991) 1;
    {\it Phys. Letters} {\bf B255} (1991) 290.
\bibitem{akv}
G. Altarelli, R. Kleiss and C. Verzegnassi (eds.),
{\it Z Physics at LEP~1},
CERN 89-08 (1989) and references quoted therein.
\bibitem{15}
D. Bardin et al., {\it Z. Physik} {\bf C44} (1989) 493;
\\
D. Bardin, W. Hollik and T. Riemann,
{\it Z. Physik} {\bf C49} (1991) 485.
\bibitem{zfitter} D. Bardin  et al.,
FORTRAN program \zf, and CERN--TH. 6443/92; based
on:~\cite{akhund,15,npb351}.
\bibitem{degrassisirl}
    G. Degrassi, S. Fanchiotti and A. Sirlin,
    {\it Nucl. Phys.} {\bf B351} (1991) 49.
\bibitem{jegerl}
F. Jegerlehner,
Physics of precision experiments with Zs,
in: A. Faessler (ed.),
{\em Prog. Part. Nucl. Phys.} (Pergamon Press, Oxford, U.K., 1991),
Vol.~27, p.~1.
\bibitem{altarjada}
G. Altarelli, R. Barbieri and S. Jadach,
{\it Nucl. Phys.} {\bf B369} (1992) 3.
\bibitem{sommi}
S. Ganguli, Tata Inst. prepr. TIFR/EHEP 91-15.
\bibitem{akhund}
 A. Akhundov, D. Bardin and  T. Riemann,
 {\it Nucl. Phys.} {\bf B276} (1986) 1.
\bibitem{buchmuwi}
G. Bhattacharyya, A. Datta, S. N. Ganguli and A. Raychaudhuri,
{\it Mod. Phys. Letters} {\bf A6} (1991) 2921; \\
W. Buchm\"uller, C. Greub and P. Minkowski,
{\it Phys. Letters} {\bf B267} (1991)395;                 \\
W. Buchm\"uller, C. Greub and H.-G. Kohrs,
{\it Nucl. Phys.} {\bf B370} (1992) 3.
\bibitem{nardi}
E. Nardi, E. Roulet and D. Tommasini,
Univ. Michigan prepr. UM--TH 92-07 (April 1992).
\bibitem{zefit}
A. Leike, S. Riemann and  T. Riemann, Univ. Munich prepr.
LMU-91/06, and FORTRAN program {\tt ZEFIT}.
\bibitem{highlumi}
E. Blucher et al. (eds.), {\it Report of the working group on high
luminosities at LEP}, CERN~91-02 (1991).
\bibitem{comblep}
The LEP Collaborations: ALEPH, DELPHI, L3 and OPAL,
{\it Phys. Letters} {\bf B276} (1992) 247.
\bibitem{layssac}
J. Layssac, F.M. Renard and C. Verzegnassi,
{\it Z. Physik} {\bf C53} (1992) 97.
\bibitem{renard2}
J. Layssac, F.M. Renard and C. Verzegnassi,
Univ. Montpellier prepr. PM/92--09 (1992),
to appear in {\it Phys. Letters} {\bf B}.
\bibitem{renorm}
P. Binetruy, S. Dawson, T.~Hinchliffe and M.~Sher,
    {\it Nucl. Phys.} {\bf B273} (1986) 501;
\\
J. Ellis, K. Enqvist, D.V. Nanopoulos and F. Zwirner,
    {\it Nucl. Phys.} {\bf B276} (1986) 14.
\bibitem{barbara}
H. Martyn et al.,      New neutral vector bosons at HERA,
contribution  to the
{\em Workshop on Physics at HERA}, DESY, Hamburg, Sept.
1991 (to appear in the proceedings).
\bibitem{cdf}
CDF Collaboration, talk by M. Gold,
in same Proc. as ref. \cite{janet},
Vol.~1, p.~388.
\end{thebibliography}
\end{document}